\documentclass[aps,pre,reprint,superscriptaddress]{revtex4-1}
\usepackage{graphicx}
\usepackage{amssymb}
\usepackage{amsmath}
\usepackage[hidelinks]{hyperref}
\usepackage{float}
\usepackage{mathtools}

\begin{document}

\title{Origin of Multiple Infection Waves in a Pandemic: Effects of Inherent Susceptibility and External Infectivity Distributions}

\author{Saumyak Mukherjee}
\affiliation{Solid State and Structural Chemisry Unit, Indian Institute of Science, Bengaluru-560012, India}
\author{Sayantan Mondal}
\affiliation{Solid State and Structural Chemisry Unit, Indian Institute of Science, Bengaluru-560012, India}
\author{Biman Bagchi}
\email[Email:]{bbagchi@iisc.ac.in}
\affiliation{Solid State and Structural Chemisry Unit, Indian Institute of Science, Bengaluru-560012, India}

\begin{abstract}
Two factors that are often ignored but could play a crucial role in the progression of an infectious disease are the distributions of inherent susceptibility ($\sigma_{inh}$) and external infectivity ($\iota_{ext}$), in a given population. While the former is determined by the immunity of an individual towards a disease, the latter depends on the duration of exposure to the infection. We model the spatio-temporal propagation of a pandemic using a generalized SIR (Susceptible-Infected-Removed) model by introducing the susceptibility and infectivity distributions to understand their combined effects, which appear to remain inadequately addressed till date. We consider the coupling between $\sigma_{inh}$ and $\iota_{ext}$ through a new Critical Infection Parameter (CIP) ($\gamma_c$). We find that the neglect of these distributions, as in the naive SIR model, results in an overestimation of the amount of infection in a population, which leads to incorrect (higher) estimates of the infections required to achieve the herd immunity threshold. Additionally, we include the effects of seeding of infection in a population by long-range migration. We solve the resulting master equations by performing Kinetic Monte Carlo Cellular Automata (KMC-CA) simulations.  Importantly, our simulations can reproduce the multiple infection peak scenario of a pandemic. The latent interactions between disease migration and the distributions of susceptibility and infectivity can render the progression a character vastly different from the naive SIR model. In particular, inclusion of these additional features renders the problem a character of a living percolating system where the disease cluster survives by migrating from region to region.
\end{abstract}

\maketitle

\section{Introduction}
\label{sec1}

Immunity of an individual and its distribution in a given population play extremely important role in the spread of any infectious disease. Yet these aspects have remained less discussed and poorly understood. Immunity is an intrinsic individual property which determines the susceptibility of an individual to a certain disease. The fraction of resilient and vulnerable population modulates the herd immunity threshold in a region.\cite{Mondal2020, Aguas2020} Hence, in order to understand the progression of an epidemic, one needs to consider the inherent distribution of susceptibility in the population.\cite{Hickson2014} Quantifying this distribution is a difficult problem because of the heterogeneity in the population in every aspect.\cite{Britton2020} An otherwise healthy population might possess low immunity towards a novel disease while a relatively unhealthy person/population could possess high immunity. The corona virus, which is responsible for the SARS-CoV-2 pandemic, seems to display some of these features.

In a recent work, we attempted to develop a statistical mechanical approach to define an immune response function called IMRF.\cite{Roy2020} We defined IMRF as the mean square fluctuation in effector T-cell (the killer cell).\cite{Roy2014} IMRF can vary from individual-to-individual and disease-to-disease. It can be quantified through standard repeated blood tests on a healthy person as mentioned in our earlier publications.\cite{Roy2020, Roy2014} It can therefore serve as a quantitative indicator, allowing us to grade the immunity of an individual according to a scale. This is more advanced than the one-shot value we obtain by a test which can be either positive or negative, and can miss the real situation.

We face further complications in modelling contagious diseases with large but slow recovery rates. These features may give rise to time dependent patterns that hinge on many factors which are hard to understand and even harder to control and model. One important factor oft ignored is the infectivity of an individual that depends on the external exposure of an individual. This is a factor to be considered in addition to the susceptibility or immunity. The external infectivity may depend on the lifestyle, travel requirement, climate etc. Thus, a person with low susceptibility (for example, a young person) can get infected if exposed to the virus for a long time, and an older person with high susceptibility can escape, owing to low exposure. Like susceptibility, this also needs to be treated as a distribution.

In addition to the above features, the time evolution of new infections also depend on migration and clustering of diseases, making the evolution both space and time dependent. These aspects are not included in the classical SIR (Susceptible-Infected-Removed) model. There have been several generalizations of the SIR model, like SAIR, SEIR etc., which include additional variables and compartments such as asymptomatic, exposed, resilient etc. However, they prove to be inadequate to address the complexities mentioned above. Very few theoretical studies have addressed the occurrence of the multiple peaks.\cite{Kaxiras2020} The endless growth and decay seem to have a pattern, at least at the intermediate times. Another interesting feature is the absence of a second peak in some countries, for example, in India. The reason for this is not yet clear. One could conjecture that this is due to some specific combination of susceptibility distribution and infectivity distribution.

The space and time dependent patterns that could evolve in a long-lasting infection are of great scientific interest and challenge, as many aspects are new and hard to predict from a theory. It is nontrivial to model the time evolution, even after taking different susceptibilities and infectiousness of individuals into account. The situation resembles a fleeting percolation scenario where infection can move around over a wide range of space and over a long period of time. Both the growth and the decay can be highly non-exponential and heterogeneous. Seasonal variations could be troublesome because even immunity of an individual could be temperature dependent. In addition, the growth also depends on the density of the population and infection in the milieu. Thus, the time and spatial evolution of new infections can show rich dynamical features. 

Emergence of multiple peaks in the COVID-19 pandemic has raised a grave concern. It has turned out to be difficult to model this phenomenon because of many factors involved. At the simplest level, one uses the following two approaches. (i) Application of time series regression analysis provides reasonably correct estimate of the new infections on a short time window, although it could become unreliable in the long run. Such regression analyses also fail to predict the occurrence of the subsequent infection peaks, after the curve flattens for the first time. (ii) Application of the standard SIR model or its variants, where one starts with a master equation describing inter-conversions between susceptible (S), infected (I) and removed (R) populations. While the second approach is based on a mathematical model, it also requires the data to be fitted into the model for correct estimation of the conversion rates. As the disease evolves, one finds that the results need to be fitted repeatedly over varying time windows for improved predictive power.

The naive SIR model consists of three coupled differential equations as described in Eq.\ref{eq1}.\cite{Skvortsov2007, Jones2009, Anderson1979} According to this model, ‘S’ may become ‘I’, and ‘I’ eventually becomes ‘R’. However ‘R’ can never become ‘S’ or ‘I’ because of the acquired immunity. The model imposes an additional constraint that at a given time t, $S(t)+I(t)+R(t)=N$, which is constant.

\begin{equation}
 \begin{split}
  \frac{dS(t)}{dt} &=-\beta S(t)I(t)\\
  \frac{dI(t)}{dt} &=\beta S(t)I(t)-\gamma I(t)\\
  \frac{dS(t)}{dt} &=\gamma I(t)
 \end{split}
 \label{eq1}
\end{equation}

Eq.\ref{eq1} describes the three coupled non-linear differential equations of the KM model, where $\beta$ is the rate of infection and $\gamma$ is the rate of removal (recovery and death). In principle, the rate constants should be time and space-dependent, that is, non-local. The naive SIR model is similar to the kinetic description of a consecutive chemical reaction.

It is important to note that the rate parameters, $\beta$ and $\gamma$ are obtained by fitting to the available data. It is often found that one set of these parameters cannot describe the full progression. One then takes the advantage of the existing data to obtain new rate parameter set valid over a certain future for predictive purpose.

From the above description, we see that the naive SIR model maybe too simple to describe the complex evolution we are witnessing in the progression of COVID-19. Certain improvements are warranted urgently in order to realize a quantitative or semi-quantitative level description. In particular, the certain inherent attributes of a population shaped by social and historical experiences and characteristics may make the progression vastly different from region to region. Thus the SIR model needs to be improved upon in the following directions.

\begin{enumerate}
 \item First, there is a need to include in the model a time dependent influx of infected population created by long distance migration. This could become a critical issue in large countries, like India and the USA where such migration is hard to control. This can lead to nucleation of the infection in a yet unaffected region, which results in multiple peaks, and much uncertainty and suffering. While control of this phenomenon is exceedingly hard, as shown by repeated resurgence of the pandemic, this surely plays a crucial role in determining the nature of pandemic propagation and evolution.
 
 \item Another difficulty lies in the uncertainty due to distribution of inherent susceptibility or immunity among the population. It is well discussed that senior citizens have lower immunity and are more susceptible or prone to infection on exposure. However, no quantitative measure in a given population is available. This alone makes a quantitative discussion difficult. 
 
 \item Yet another complex issue is the variation in the infectivity of a person with a given susceptibility, due to exposure. A part of the population undergoes repeated exposures through travels to offices and attending schools and colleges or other areas of public gathering. These exposed people are more prone to infection than the ones who either stay at home or work at isolated environments like fields as in agriculture.
\end{enumerate}

The effects of susceptibility distribution have been treated in several recent papers. To the best of our knowledge, Hickson and Roberts first addressed this issue of the hidden role that a susceptibility distribution can play in the progression of an epidemic.\cite{Hickson2014} More recently, Britton et al. and Aguas et al. have explicitly treated the effects of this distribution in the context of COVID-19.\cite{Aguas2020, Britton2020} In the latter case, the overriding concern about the vulnerability of the older cross-section of the population has driven these studies. However, even in the age group of 11-65 years, there could be a large number of people who have low susceptibility or higher immunity.

In our studies, we include the distributions of inherent susceptibility ($\sigma_{inh}$) and external infectivity ($\iota_{ext}$) in a population. The distributions can vary from population to population and from region to region. The values of inherent susceptibility and external infectivity together determine infection. The coupling between the effects of the two distributions is included by considering a coupled parameter which is given by

\begin{equation}
 \gamma = \sigma_{inh} \times \iota_{ext}
 \label{eq2}
\end{equation}

A susceptible person in contact with an infective will get infected if $\gamma$ is greater that the Critical Infection Parameter (CIP) denoted by $\gamma_c$. There are certain limiting conditions, which the CIP must satisfy. A person, even if highly susceptible, may not be prone to infection, if he/she is isolated (home quarantine, for example) and does not come in contact with a highly infective individual. On the other hand, a person with lower susceptibly may get infected easily if he frequents regions surrounded by persons with high infectivity. These limiting conditions rule out the possibility of $\gamma$ being defined by an additive rule between $\sigma_{inh}$ and $\iota_{ext}$. Hence, we use a multiplicative definition, which appears to satisfy the aforementioned limits. However, defining $\gamma$ using a proper functional form is nontrivial and can only be obtained phenomenologically by comparing with the real-world scenario.

In some countries and locations, there is certain homogenization of the exposure because young people often live with older population, leading to a close proximity of resilient and vulnerable sections of the population. However, in other countries this might not be the case. The more affluent the society, the more segregated the population becomes. Thus, the infection could remain localized among a section of population. Thus, we begin to see that even when a part of the population is more susceptible, they could be less exposed and less prone to infection. This could affect infection rate later. The infectivity may also depend on the seasons, with winter being more infective due to increase in the indoor activity. Thus, in the language of physics, we need a multidimensional approach. We need to consider both susceptibility and infectivity together in order to ascertain the probability of infection in a given individual.

To summarize the discussion above, in this study we have included these two (new and scantly addressed) factors, the inherent susceptibility (which is connected to immunity) and the external infectivity (which is connected to the exposure time to infection), as distributions. Next, we use a multiplicative rule (as being the simplest) to determine the probability of infection (Eq.\ref{eq2}). Thus, the probability of infection of an individual depends on the coupled effects of duration of exposure and immunity.

It is clear that the understanding and also the usefulness of the concept of herd immunity need modification in the presence of susceptibility distribution. In the presence of a sizeable fraction of population with low susceptibility, the percentage of people that needs to develop anti-body by infection could decrease significantly, as has already been noted by Aguas et al.\cite{Aguas2020}

Another new aspect is the inclusion of infection probability via long distance migration. This is often the cause of multiple peak disease progression. This is to be combined with seasonal variations which, as mentioned above, give rise to an increase in the infectivity of a person. 

The objective of this study is to explore these different aspects through a generalized SIR model. Our new equations are not just non-linear. They are also non-local and stochastic because rate parameters are derived from a distribution. We solve the problem through cellular automata simulations and explore effects of variations in parameters on the multi-peak infection distribution that unfolds with time.
 
The main results of the work is that the presence of the distributions can significantly alter the time dependent progression of infection from the predictions of the simple or naive SIR model. In particular, the presence of distributions could reduce the infection peak height. Secondly, the distributions combined with the migration induced “injection” of the disease can give rise to multiple peaks as has been indeed observed. This in itself is not surprising, except the present generalized formalism might enable a more quantitative description than usually employed.

The presence of a significant fraction of low susceptibility people can lower the threshold of herd immunity. This has already been observed by Aguas \textit{et al}.\cite{Aguas2020} These low susceptibility fraction of people can be excluded from the total population of susceptible. This rather simple aspect can however be included only through a distribution, as has been done in the present work. The situation with highly susceptible but less infective population is a delicate matter to handle because this fraction always remains vulnerable. Treatment of such populace also needs an explicit use of distributions.

The rest of the paper is organized as follows. In Section \ref{sec2}, we lay out the theoretical formulation used in this study. Next comes Section \ref{sec3} which deals with the spatio-temporal dependence of the different compartments of the population and the associated differential equations that define their dynamics. This is followed by Section \ref{sec4}, which includes a thorough description of the Kinetic Monte Carlo Cellular Automata (KMC-CA) simulation scheme, which is used to study the infection dynamics. In Section \ref{sec5}, we present the results obtained from the simulations and discuss the consequent implications and inferences. Starting from a nucleation of infection, the disease spreads throughout the community via a percolation network. We obtain the multiple peak nature like a real-world pandemic (for example, Spanish flu, COVID-19, etc.). Finally, we conclude the work in Section \ref{sec6}.

\section{Theoretical Formulation: Distribution Based Model}
\label{sec2}

\begin{figure}[ht]
 \centering
 \includegraphics[width=3in,keepaspectratio=true]{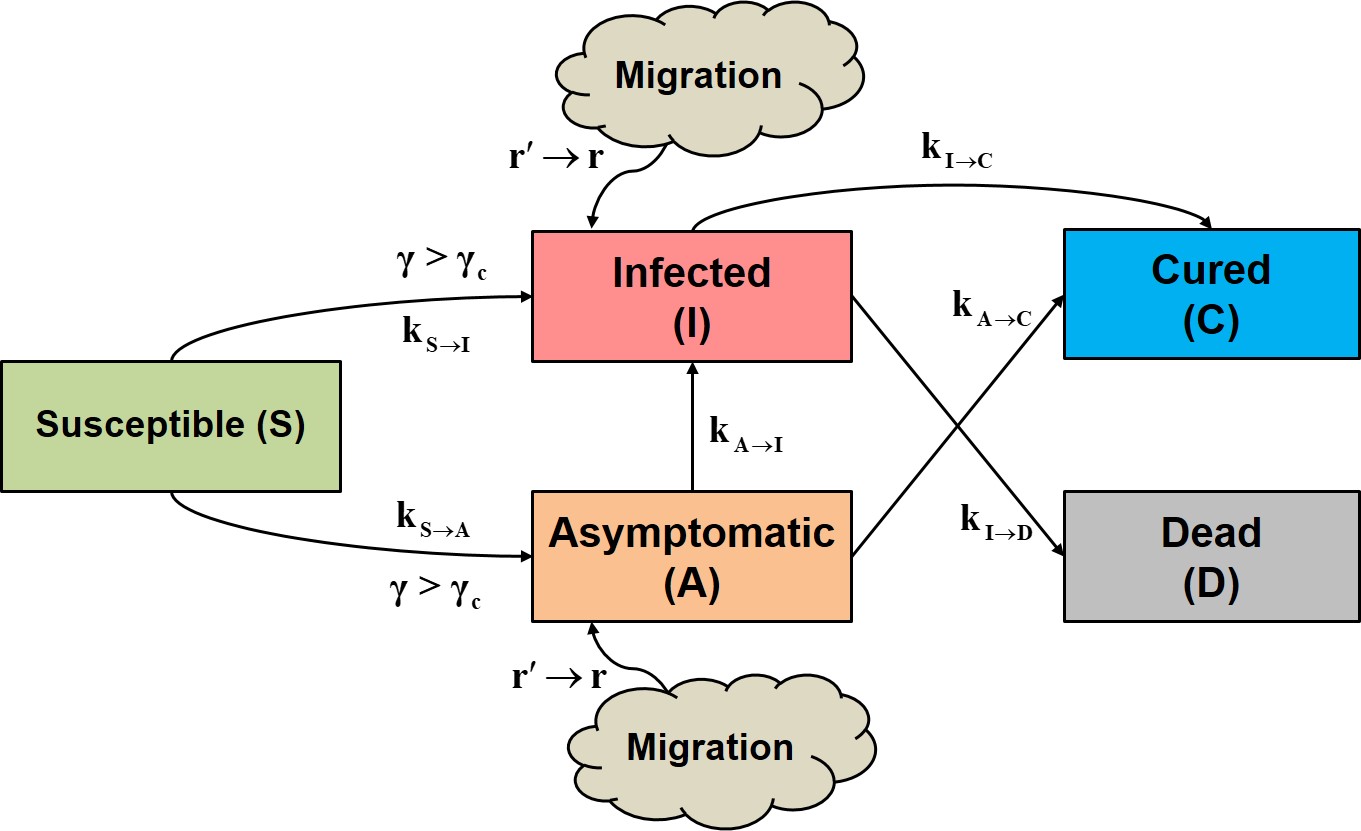}
 \caption{Schematic representation of our pandemic model where the susceptible (S) population, after being exposed to the virus, can either become Infected (I, with symptom) or Asymptomatic (A). For an uninfected individual, if the product ($\gamma$) of the inherent susceptibility and external infectivity reaches a certain critical value ($\gamma_c$) (Critical Infection Parameter, CIP) then the person becomes infected. The I compartment can either become Cured (C) or Dead (D). On the other hand, a fraction of the A compartment might develop symptom and become (I). The other fraction gets naturally cured. In addition to these, we consider a random seeding event to incorporate the effect of spatial migration of infection into the simulated locality from outside. This makes the total density globally conserved, but not locally conserved. The rate constants associated with these processes are written on/below the corresponding arrows.}
 \label{fig1}
\end{figure}

The central quantities in our discussion are space and time dependent densities of susceptible (S), infected (I), asymptomatic (A), cured (C) and dead (D) persons. In Fig. \ref{fig1}, we schematically represent the complex network that is involved in such a disease transmission. The model is inspired from the celebrated Susceptible-Infected-Removed (SIR) model proposed long ago by Kermack and McKendrick.\cite{Kermack1927, Diekmann1995, Diekmann2000, Siettos2013} It is noteworthy that many recent studies have employed this model and its variants in the context of the COVID-19 pandemic.\cite{Mukherjee2020, Wangping2020, Sirakoulis2000, Petropoulos2020, Kucharski2020}

According to the present model, a susceptible (S) individual (with an inherent susceptibility index $\sigma_{inh}$) can either become infected (I) with symptoms or become asymptomatic (A), by getting exposed to an infected or asymptomatic individual (with an external infectivity index, ($\iota_{ext}$ or $\iota_{ext}^A$ respectively). When the product of the susceptibility and infectivity indices ($\gamma = \sigma_{inh} \times \iota_{ext}$) reaches a threshold value ($\gamma_c$), the susceptible individual gets infected.
The I compartment of the population might be cured (C) or dead (D) with the passage of time. There is also a probability that a fraction of the A category develops symptoms and becomes I. The rest of A becomes cured without any fatality. In order to incorporate the effect of spatial migration of infection, we consider a random seeding event that increases the infected population by $\delta I$.

The densities of these aforementioned variables all depend on space and time. We note that due to the inclusion of migration the density of the system is not locally conserved, but globally conserved. Hence, there are five density terms with the following global conservation constraint

\begin{equation}
 \begin{split}
 &\rho_S(\mathbf{r},t)+\rho_A(\mathbf{r},t)+\rho_I(\mathbf{r},t)\\
 &+\rho_C(\mathbf{r},t)+\rho_D(\mathbf{r},t)=\rho(\mathbf{r},t)
 \end{split}
 \label{eq3}
\end{equation}

It is important to note that the total density $\rho(\mathbf{r},t)$ itself is a local variable and introduces a degree of heterogeneity in the overall population. This could vary from region to region, like from a city dwelling to a village or rural surrounding. Even within a city, there could be vastly different densities, like those in slums and affluent localities. The density can vary more than one order of magnitude.

A surprising initial observation is that susceptibility is different not just between different age groups but also between different localities. In India, for example, the susceptibility appears to be smaller in slums.\cite{Hasan2020, Brotherhood2020} If indeed true, one could explain invoking immunity, but the origin is not clear. 

It is clear from the above discussion that any predictive theory needs to include a large number of parameters and that neither the time series expansion method nor the simple SIR model can possibly capture the complex dynamics. Given that we need at least a semi-quantitative understanding it is perhaps prudent to attempt a theory of intermediate complexity. Of particular importance are the following. (i) A distribution of inherent susceptibility in population, (ii) a distribution of external infectivity which is dependent on seasonal changes and time of exposure, (iii) local population density, and (iv) long range disease transfer by travel or migration.

As discussed before, we introduce two parameters, namely, the inherent susceptibility index ($\sigma_{inh}$) and the external infectivity index ($\iota_{ext}$) respectively. These two parameters together control the probability of a susceptible individual to get infected from an infected individual. For example, infected (or asymptomatic) individuals who wear mask, practice good respiratory hygiene, and avoid crowds possess a low value of $\iota_{ext}$. On the other hand, susceptible individuals with high intrinsic immunity possess a low $\sigma_{inh}$. Hence, the quantity ($\gamma = \sigma_{inh} \times \iota_{ext}$) must be above a certain threshold, $\gamma_c$, for the infection to spread. These two values are randomly sampled from a pre-existing distribution.  is another parameter that we introduce to scale the strength of infection in the case of A category people.

\section{Equations of Motion for Densities}
\label{sec3}

We now present the equations for the time dependence of the dynamical variables as mentioned above [in Eq.\ref{eq3}]. The density terms should ideally also be dependent on $\sigma_{inh}$ and $\iota_{ext}$, and needs to be written as $\rho(\mathbf{r},t|\sigma_{inh},\iota_{ext})$. However, for simplicity we drop the indices corresponding to susceptibility and infectivity. Therefore, we can write the following coupled non-local differential equations where $k_{\alpha\rightarrow\beta}$ denotes the rate-constant of transition from compartment $\alpha$ to compartment $\beta$. 

\begin{widetext}

\begin{equation}
 \begin{split}
  \frac{\partial \rho_S({\bf r},t)}{\partial t} = &-\rho_S({\bf r},t)\int d{\bf r'}\sum_{\sigma_{inh}, \iota_{ext}} k_{S \to I}({\bf r}-{\bf r'},t|\sigma_{inh}, \iota_{ext})\rho_I({\bf r}-{\bf r'},t)\\
  &-\rho_S({\bf r},t)\int d{\bf r'}\sum_{\sigma_{inh}, \iota_{ext}^A} k_{S \to A}({\bf r}-{\bf r'},t|\sigma_{inh}, \iota_{ext}^A)\rho_A({\bf r}-{\bf r'},t)
 \end{split}
 \label{eq4}
\end{equation}

\begin{equation}
 \frac{\partial \rho_A({\bf r},t)}{\partial t} = \rho_S({\bf r},t)\int d{\bf r'}\sum_{\sigma_{inh}, \iota_{ext}^A} k_{S \to A}({\bf r}-{\bf r'},t|\sigma_{inh}, \iota_{ext}^A)\rho_A({\bf r}-{\bf r'},t)-\rho_A({\bf r},t)(k_{A \to I}+k_{A \to C})
 \label{eq5}
\end{equation}

\begin{equation}
 \begin{split}
  \frac{\partial \rho_I({\bf r},t)}{\partial t} &= \rho_S({\bf r},t)\int d{\bf r'}\sum_{\sigma_{inh}, \iota_{ext}} k_{S \to I}({\bf r}-{\bf r'},t|\sigma_{inh}, \iota_{ext}) \rho_I({\bf r}-{\bf r'},t)\\
  &-\rho_I({\bf r},t)\int d{\bf r'}\Big[k_{I\to C}({\bf r}-{\bf r'})+k_{I \to D}({\bf r}-{\bf r'})\Big]+k_{A \to I}\rho_A({\bf r},t)\int d{\bf r'} T({\bf r'}\to{\bf r})\rho_I({\bf r},t)
 \end{split}
 \label{eq6}
\end{equation}

\begin{equation}
 \frac{\partial \rho_C({\bf r},t)}{\partial t} = k_{A\to C}\rho_A({\bf r},t)+\rho_I({\bf r},t)\int d{\bf r'}k_{I\to C}({\bf r}-{\bf r'})
 \label{eq7}
\end{equation}

\begin{equation}
 \frac{\partial \rho_C({\bf r},t)}{\partial t} = \rho_I({\bf r},t)\int d{\bf r'}k_{I\to D}({\bf r}-{\bf r'})
 \label{eq8}
\end{equation}

\begin{equation}
 k_{S\to I}({\bf r}-{\bf r'},t|\sigma_{inh},\iota_{ext}) = k_{S\to I}({\bf r}-{\bf r'},t)H(|{\bf r}-{\bf r'}|\le r_0)H(\gamma\ge \gamma_c)\Big[\frac{4}{3}\pi r_0^3\Big]
 \label{eq0}
\end{equation}

\end{widetext}

$\gamma$ is given by Eq.\ref{eq2}. Eq.\ref{eq0} is true for $k_{S\to A}({\bf r}-{\bf r'},t|\sigma_{inh}, \iota_{ext}^A)$ as well. $H$ denotes Heaviside functions. Hence, for the rate constants to possess non-zero values, the distance between S and I (or A) must be less than or equal to a cut-off distance $r_0$. In addition, $\gamma$ must be greater than a critical value $\gamma_c$. $T({\bf r'}\to{\bf r})$ is a transfer term that allows infections from ${\bf r'}$ to come to ${\bf r}$, and $\iota_{ext}^A = \iota_{ext}\chi$. We note that, some of the rate constants that are disease specific, for example, the speed of recovery and mortality rate, are assumed to be independent of time but dependent on space as the healthcare facilities are spatially heterogeneous. On the other hand, the rate constants associated with transition from A to some other compartment, are naturally independent of both time and space. 

These equations possess a striking resemblance with the chemical reaction kinetics network theory or coupled parallel chemical reactions. In the next section we shall discuss the method of solving these equations numerically with the help of kinetic Monte Carlo cellular automata simulations.

\section{Solution by Kinetic Monte Carlo Cellular Automata (KMC-CA) Simulations}
\label{sec4}

The spatio-temporally resolved differential equations, that define the dynamics of the multiple compartments of population during a pandemic, are nontrivial and cannot be readily solved analytically. However, a numerical approach can be perceived to understand this disease dynamics in terms of Kinetic Monte Carlo Cellular Automata (KMC-CA) simulations.

Cellular automata is a popular technique to study physical processes like chemical reactions, wildfire propagation, traffic dynamics, phase transitions, pattern formation etc.\cite{Hollingsworth2004, Seybold1998, Wolfram1983, Bartolozzi2004, Vannozzi2006, Almeida2011, Weimar2002, Kier2005} This technique has also been used to study the progress of epidemics.\cite{Pfeifer2008, Zhong2009, Athithan2014, White2009, White2007, Tiwari2020, Monteiro2020} We perform KMC-CA simulations of the present model with several parameters and factors that mimic the spread of an infectious disease into a population of susceptible individuals. As mentioned in the previous section, the model posits that during an ongoing pandemic, at any point of time, society consists of 5 major types of individuals, namely, susceptibles (S), asymptomatic infectives (A), symptomatic infectives (I), cured (C) and dead (D). The salient features of the present KMC-CA model are discussed below.

\begin{enumerate}
 \item We start with a 2-dimensional area denoted by a matrix of $N_X \times N_Y$ cells. A given fraction of the area is covered by S and I individuals. The positions of these individuals are assigned randomly. There are no A, C or D in this initial frame. This gives the initial configuration of the population.
 
 \item A person moves by randomly choosing the direction among the 8 available grids adjacent to the present cell. These movements may be biased or restricted for reasons described below (points \ref{pt7} and \ref{pt9}). The time taken by a person to move to the next neighbouring cell serves as the unit of time in our simulation.
 
 \item Each individual is assigned with an inherent susceptibility index ($\sigma_{inh}$) and an external infectivity index ($\iota_{ext}$), both of which are sampled from given distributions. We use three distributions for the susceptibility, namely, Gaussian, bimodal Gaussian and uniform. Infectivity is sampled from a Gaussian distribution. The value of these indices lie between 0 and 1. We assume that the values of $\sigma_{inh}$ and $\iota_{ext}$ remain constant throughout the lifetime of the individual. The value of $\sigma_{inh}$ quantifies the immunity of an individual. $\iota_{ext}$, on the other hand determines how prone an infected person is to spread the infection. It is assumed that an S can get infected if the following two criteria are fulfilled:
 
  \begin{itemize}
   \item The S should be in either of the 8 cells surrounding an I (Moore neighbourhood \cite{Sirakoulis2000, White2007, Fu2003})
   
   \item The value of $\gamma = \sigma_{inh}\times\iota_{ext}$ for this pair of S and I should be greater than a given critical value, the Critical Infection Parameter (CIP) ($\gamma_c$). $\gamma$ is proportional to the rate constant $k_{S\to I}$ in Eqs. \ref{eq4} and \ref{eq6}. It is a coupled parameter that accounts of the extent to which the concerned persons are taking care of personal and social protection, like wearing a mask, maintaining physical distance and so on. 
  \end{itemize}
  
 This protocol for the spread of infection is also true for the interaction between S and A. Details about A is given in point \ref{pt5} below.

 \item From the initial step, each individual is assigned an age according to a given distribution. In our present simulations, we use the age distribution given in Table \ref{tab1}(\href{https://www.indexmundi.com/india/demographics_profile.html}{Reference}).

 \begin{table}[ht]
\caption{Age distribution in India in 2019.}
 \centering
 \begin{tabular}{|c|c|}
  \hline
  Age window (years) & Percentage of population\\
  \hline
  \hline
  0-14	& 26.98\\
  \hline
  15-24	& 17.79\\
  \hline
  25-54	& 41.24\\
  \hline
  55-64	& 7.60\\
  \hline
  65+	& 6.39\\
  \hline
  \end{tabular}
 \label{tab1}
\end{table}

 Note that, any desired age distribution can be used in the simulation, based on the demography of the geographical area under consideration. Two ages, Age-1 and Age-2 are parameterized in the simulations. An individual is categorized as resilient (Res) if Age-1 $<$ age $<$ Age-2. Otherwise, the individual is vulnerable (Vul). Immunity of a person from corona virus shows dependence on age.\cite{Singh2020} If infected, the vulnerables have a lesser probability of recovery as compared to the resilients. In context of the SARS-CoV-2 infection it is seen that the elderly and the infants are more vulnerable. Age-1 and Age-2 are chosen accordingly.
 
 \item \label{pt5} A major problem in controlling the COVID-19 pandemic is the emergence of asymptomatic carriers who act as silent spreaders of the virus.\cite{Long2020, Zhang2020a} Reports show that about 40 \% of infected individuals are asymptomatic.\cite{Nishiura2020, Oran2020} In our KMC-CA simulation, we provide probability ($P_{asym}$), which decides whether an infected individual will be asymptomatic (A) or not. Initially, a random asymptomatic index ($\iota_{ext}^A$) is assigned to each individual which is activated on getting infected. This variable is compared with $P_{asym}$ to determine the fate (A or I) of the infected person. The strength of infectivity of an A is different from that of I. For SARS-CoV-2 A is found to be less infective than I.\cite{McEvoy2020} However, some studies also suggest that these two categories of infected people may show similar disease transmissibility.\cite{He2020} This can be modulated by a factor $\chi$ according to Eq.\ref{eq9}.
 
 \begin{equation}
  \iota_{ext}^A = \iota_{ext}\chi
  \label{eq9}
 \end{equation}
 
 where, the value of $\chi$ generally lies between 0 and 1. A certain fraction of asymptomatics (A) may develop symptoms, after a certain time of getting infected. The rate of this conversion is given by $k_{A \to I}$ in Eqs. \ref{eq5} and \ref{eq6}. In the KMC-CA simulation, it is modulated by a probability ($P_{A \to I}$).

 \item In each step of the simulation, an I can either remain infected of recover or die. This is determined by the following probabilities:
 
 \begin{itemize}
  \item $P_I$ : This determines whether the concerned I remains infected or not. This is proportional to the incubation period of the virus and the time period for which a person remains infected. We find that for a disease like SARS-CoV-2 to persist in a society this probability needs to be very high ($P_I > 0.9$).
  
  \item $P_{I \to C}^{Res}$: This gives the probability of recovery of a resilient I.
  \item $P_{I \to C}^{Vul}$: This gives the probability of recovery of a vulnerable I.
 \end{itemize}
 
 We consider $P_{I \to C}^{Res} > P_{I \to C}^{Vul}$.\cite{Mondal2020} Asymptomatic infectives (A) always show complete recovery, i.e. $P_{A \to C} = 1$ irrespective of whether the person is resilient or vulnerable.
 
 Once cured, the person generally becomes immune to further infection as the disease specific antibodies are generated. The period of immunity may extend to as long as 6 months or more.\cite{Dan2020} After that, reinfection may occur. Hence a probability is introduced ($P_{C \to S}$) which determines the conversion of C to S so that the person again becomes susceptible to infection.
 
 \item \label{pt7} An important manoeuvre employed by most governments to control the outbreak of COVID-19 is a lockdown of the citizens. While the infected people are either in home quarantine or in hospitals, others are advised to stay indoors as a lockdown measure. To account for this scenario in our KMC-CA simulation, we introduce two probabilities, $P_Q$ and $P_{LD}$, which restrict the movements of I and S respectively. It is important to note that since the asymptomatics (A) remain undetected, their movements are like those of S. This adds to the rate of disease spread in a population. It should be noted that cured individuals (C) need not follow these rules since they are immune to further infection, and also cannot spread the disease.
 
 However, several reports show that some fraction of people fail to abide by the lockdown/quarantine norms. As a psychological issue, people are often found to relax these norms, particularly after a certain period from the commencement of the lockdown. Also, the incipient problems like economic downfall (among others) as a consequent of national or regional lockdowns, the Governments are forced to relax the rules. For example, India has followed step-wise “unlock” procedures to allow normal movement of its citizens. Hence, while quarantine of I remains strict, lockdown measures are lifted. 

 We use a switch parameter that allows us to either employ or neglect the above psychological factor in a simulation run. If neglected, $P_{LD}(t) = P_{LD}(0)$ throughout the simulation.
 
 \item A major contributor to spread of COVID-19 is the migration of the disease, carried by people travelling from one place to the other inside a country or even abroad. This is advocated in our simulation by randomly introducing infected individuals with on the area of our simulated society. Note that while the total number of people (including the deceased) does not remain conserved in a locality, a global conservation of population is inherent. However, the latter scenario is out of the scope of our simulation. It should be noted that the present KMC-CA simulation mimics only locality during a pandemic and does not include the global (world-wide or country-wide) outbreak scenario. The inherent heterogeneity in the nature of disease spread introduced by several geographic, atmospheric, demographic, political and other factors make it almost impossible to simulate a global outbreak. 
 
 \item \label{pt9} Local gatherings in markets, clubs, gymnasiums etc. can accelerate the process of infection significantly. We introduce gathering spots at random locations in our simulated society. The number of such spots is parameterized. We define two age limits. Only the individuals within these limits can participate in the gatherings. This approximation is validated by the fact that infants or very old people do not generally go to markets, gymnasiums, clubs etc. Even, within these age limits, a probability defines whether the person will go to a gathering or not. 
 
 To make the scenario simple, we consider that a person whose movement is biased, can go to the nearest gathering point only. We note that a person moving towards a gathering spot executes a biased random walk. Once within a defined spatial limit of a gathering point, the individual spends some time in that region, after which the bias is lifted from his/her movement, so that free movement is resumed.

 This is a feature included in our simulation, but not been used in this work. Hence, in the present work, the number of gathering points have been considered as 0.
\end{enumerate}

\section{Results and Discussion}
\label{sec5}
 
Let us emphasize at the very outset that the progression of an epidemic has been found to be strongly dependent on the characteristics of the distribution, and also on our choice of the critical infection parameter (CIP). The main observation is that the number of infection decreases from the prediction of the naive SIR model because of the presence of the distributions. We next present the results of our simulations.

\subsection{Disease Percolation Network}
\label{subsec1}

As already mentioned, direct solution of Eqs. \ref{eq4} to \ref{eq8} is extremely nontrivial. Hence we use Kinetic Monte Carlo Cellular Automata (KMC-CA) technique to simulate the system that can be exactly described by these equations. The infection starts from a single person, often termed as “patient zero” and spreads throughout the entire community very fast. For an infectious disease like corona virus, the mode of transport is person-to-person contact, via droplet exchange. This process is aggravated if the infected person is not detected at an early stage and quarantined. In such a scenario, the infectives can move around and spread the disease. This is also true for asymptomatics. 

\begin{figure}[ht]
 \centering
 \includegraphics[width=3in,keepaspectratio=true]{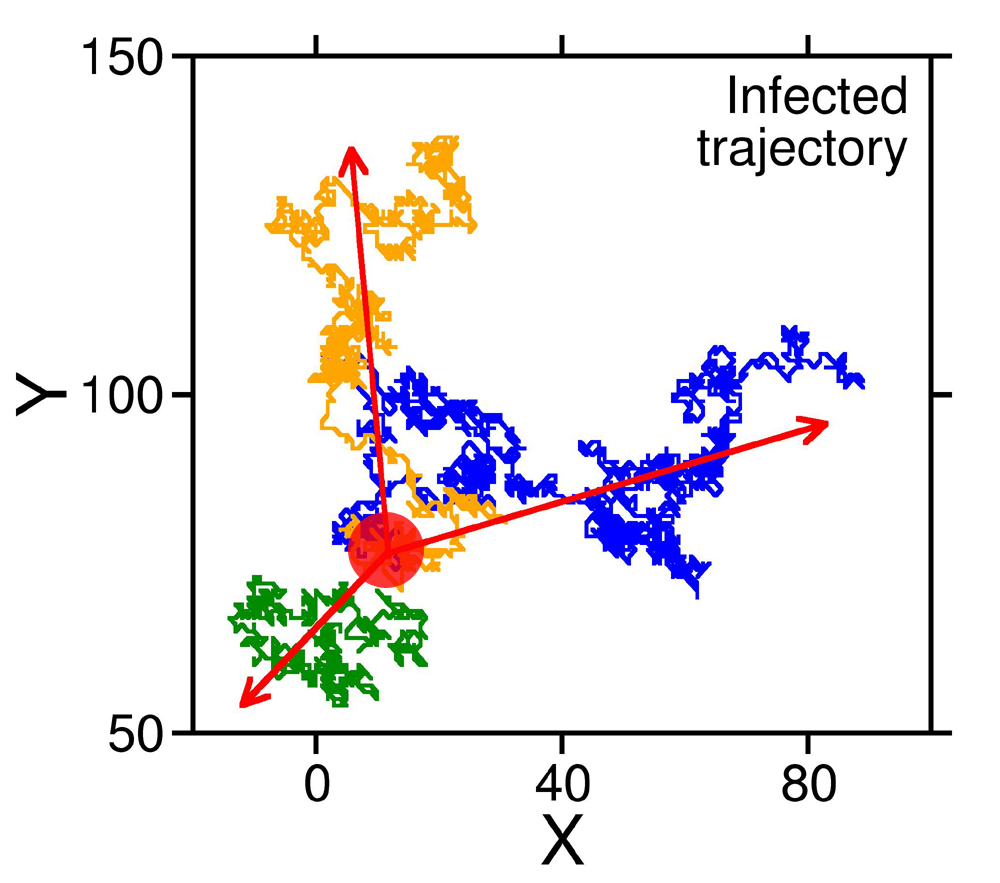}
 \caption{Trajectory of three infected individuals originating at the same point (denoted by the red circle). Beginning from a single nucleus, the infection spreads isotropically forming a percolation network. Such diffusion of infection into the population can be checked by quarantine measures, however, the infection may still infiltrate the society via asymptomatic individuals, who cannot be detected.}
 \label{fig2}
\end{figure}

From KMC-CA simulations, we can monitor the movement of these infected people. When the quarantine probability ($P_Q$) is 0 or low, or the infectives in question are asymptomatic (thus moving like susceptibles), they can fan out in all directions, carrying the disease. The trajectories of three such infectives are shown in Fig. \ref{fig2}. They originate from approximately the same point (shown by a red circle) and move out isotropically into the population of susceptibles. Implementation of strict quarantine measures can stop these movements. However, asymptomatic people can still spread the disease in the susceptible population.

\begin{figure}[ht]
 \centering
 \includegraphics[width=3in,keepaspectratio=true]{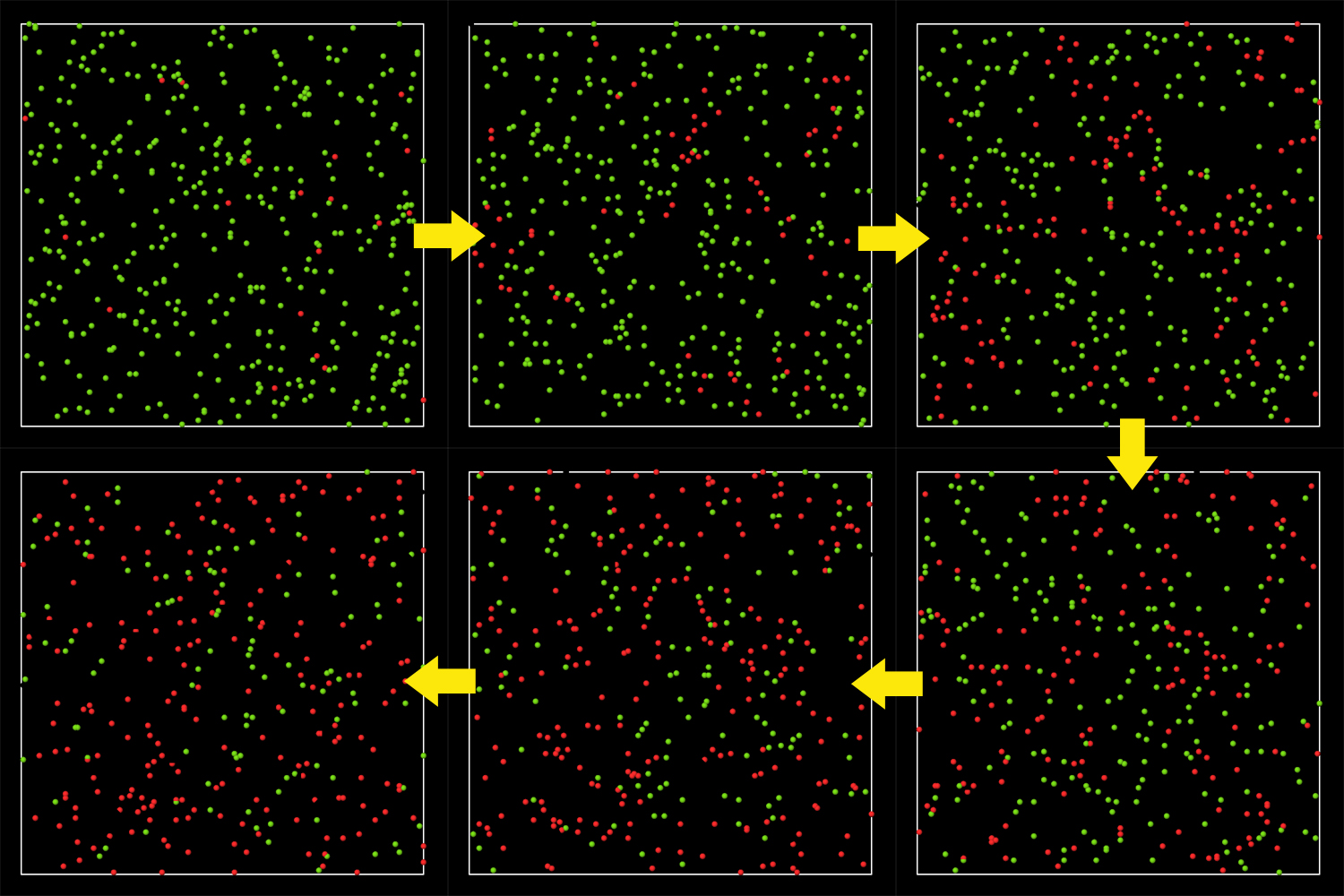}
 \caption{Spatio-temporal propagation of an infectious disease. These are snapshots from KMC-CA simulation explained in the previous section. The green and red coloured dots represent Susceptible (S) and Infected (I) individuals in the population. An infectious disease can diffuse into the susceptible population very fast, depending on the inherent susceptibility and external infectivity of the people. For clarity of representation of the percolation of disease, the recovered individuals are not shown here.}
 \label{fig3}
\end{figure}

Such movement results in a percolation network in the population, as depicted in Fig. \ref{fig3}. In this figure, the colours green and red represent susceptibles and infectives respectively. This clearly shows the infection map, and how the situation can give rise to a pandemic very quickly. It is interesting to note that the propagating network exhibits a fractal character, with a fractal dimension $d_f$ considerably less than 2. 

Hence, from a physical point of view, the pandemic starts with a nucleation of infection, which grows in an isotropic nature that results in a percolation network of infection, finally leading to a phase separation between susceptibles, infectives and recovered people.

\subsection{Effect of Susceptibility Distributions}
\label{subsec2}

The immunity of different population towards a disease is heterogeneous in a given population. The distribution of susceptibility represents this heterogeneity. Clearly, without considering this distribution, it is impossible for any model to predict the proper outcome of a pandemic.

\begin{figure}[ht]
 \centering
 \includegraphics[width=3.3in,keepaspectratio=true]{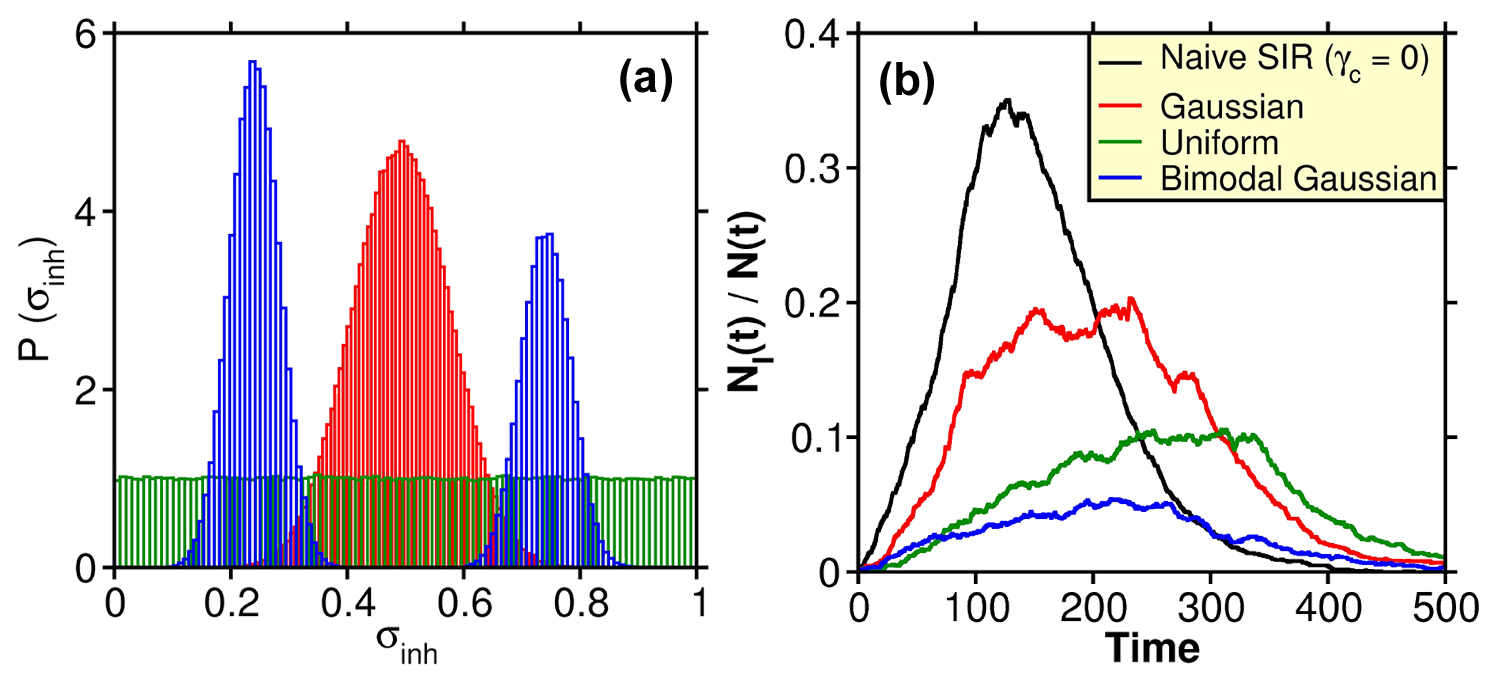}
 \caption{(a) Distribution of susceptibility in a population. In this work we have considered three types of distributions, namely, Gaussian (red), uniform (green) and bimodal Gaussian (blue). The effect of inclusion of the susceptibility distribution on the temporal evolution of infection curve is shown in (b). The ordinate represents the fraction of total population infected ($N_I(t)$ and $N(t)$ are the number of infected people and total number of people at any given time t). The unit of time is given by the time taken by a person to move from the present cell to a neighbouring cell. Naive SIR denotes the classical SIR model, where a susceptible person gets infected as soon as he/she comes in contact with an infected individual. This is obtained by setting the Critical Infection Parameter (CIP) $\gamma_c = 0$. It shows that the SIR model overestimates the amount of infection in a given population, which also results in erroneous evaluation of herd immunity threshold.}
 \label{fig4}
\end{figure}

However, quantification of the susceptibility of individuals in a population is a daunting task. There is no well-established scale of susceptibility that can be used to generate such distributions. Hence, we use certain model distributions in our simulation to investigate their effects on the time evolution of a pandemic. It is to be noted that we do not address any real world population quantitatively. Our work is aimed at developing a model, which is a significant development over the classical SIR scheme. The distributions used in these work are: (i) Gaussian (i) Uniform and (iii) Bimodal Gaussian. These are shown in Fig. \ref{fig4}a. Hickson et al. have used similar distributions.(3)

A Gaussian nature is manifested in most natural phenomena. Hence, this might also be true for the distribution of susceptibility in a given population. A bimodal Gaussian distribution can result from widely different living standards and the resultant immunity variations in a region; for example, the difference between slum dwellers and city dwellers in an urban milieu. 

We run multiple KMC-CA simulations using these distributions of susceptibility. The resultant time evolution of the fraction of infectives (I) is shown in Fig. \ref{fig4}b. We have shown four representative trajectories. For these simulations, we have switched of the random seeding of infection (that represents long range migration) to avoid complications. This is dealt with in Fig. \ref{fig7}.

In Fig. \ref{fig4}b, the black curve represents the naive SIR model, without the presence of any distribution, such that a susceptible person becomes infected as soon as an infective is present in the neighbouring cell. This is advocated by setting the Critical Infection Parameter (CIP) $\gamma_c = 0$. In the absence of the effect of distributions and long-range migration, this represents the classical SIR model. Comparison of the infection curves in Fig. \ref{fig4}b makes it clear that distributions of susceptibility is pivotal to the proper estimation of infection prevalence in a community. In absence of this consideration, the model clearly predicts a significantly higher number of infections. 

For the three simulations with the susceptibility distributions, the value of CIP was fixed at $\gamma_c = 0.25$. While Gaussian distribution gives a lower fraction of infection as compared to the “no distribution” scenario, a bimodal Gaussian results in the lowest peak height. Uniform distribution gives intermediate result. 

\begin{figure}[ht]
 \centering
 \includegraphics[width=3in,keepaspectratio=true]{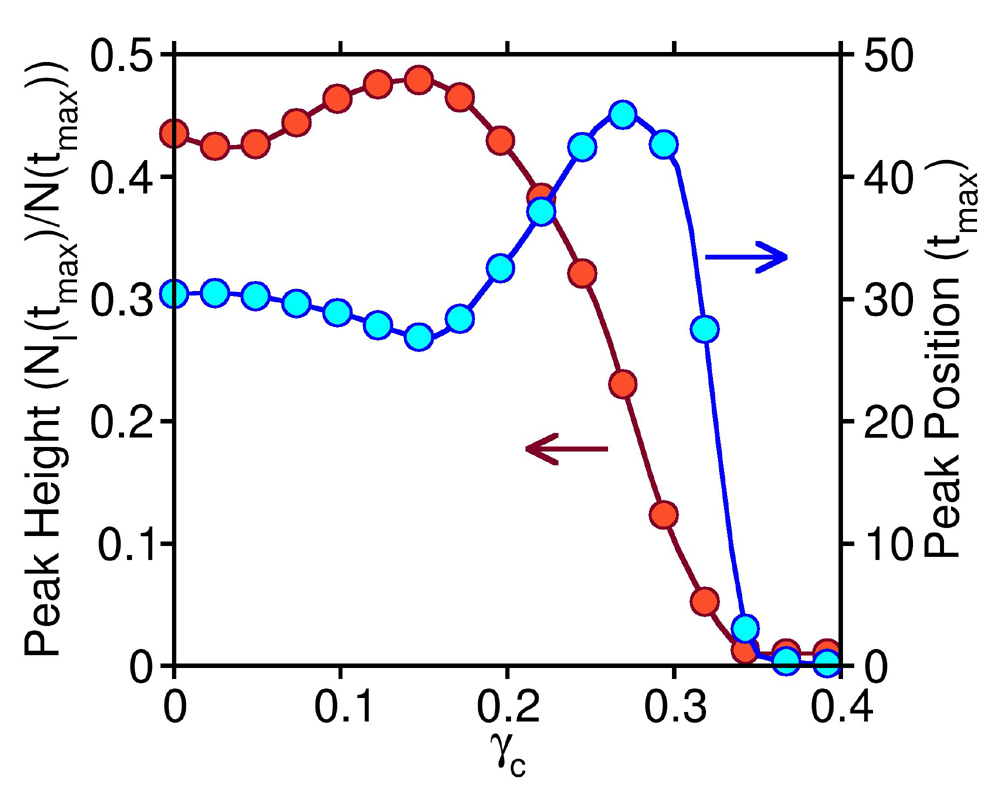}
 \caption{The change in the infection peak height ($N_I(t_{max})/N(t_{max})$) (red, left ordinate) and position ($t_{max}$) (blue, right ordinate) as a function of the Critical Infection Parameter ($\gamma_c$). The change is highly nonlinear. There is minimal change between $\gamma_c = 0$ and $\gamma_c = 0.15$. Beyond $\gamma_c = 0.3$, both peak height and position become negligible, which denotes that susceptibles cannot get infected.}
 \label{fig5}
\end{figure}

In Fig. \ref{fig5} we investigate the sensitivity of the infection peak height (red, left ordinate) and position (blue, right ordinate) of the fraction of infectives to the value of the CIP ($\gamma_c$). Each simulation is performed with the Gaussian distribution of susceptibility. We find that both peak height and position show minimal change between $\gamma_c = 0$ and $\gamma_c = 0.15$, after which the variation becomes highly nonlinear. For $\gamma_c > 0.3$, the values of height and position become negligible, denoting the absence of disease spread in the population. 

The dependence of the infection propagation the CIP has severe consequences. According to our model, infection can only spread if neighbouring susceptible and infective persons have value of $\gamma = \sigma_{inh} \times \iota_{ext} > \gamma_c$. Hence both $\sigma_{inh}$ and $ \iota_{ext}$ have to be sufficiently high for the susceptible person to get infected. Consequently, if a highly susceptible individual comes in contact with a person having low susceptibility index (and vice versa), the infection will not propagate. This is a practical situation, since wearing masks, washing hands, using sanitizers, staying indoors and other safety protocols can significantly reduce the number of infections in a population.

\subsection{Origin of Multiple Infection Peaks}
\label{subsec3}

In general, solution of the classical SIR model shows a single maximum in the temporal evolution of the number of infectives. However, in a real world pandemic, the nature of the infection is not so simple. For example, Spanish flu in 1918 was characterized by 3 peaks of mortality and infection,\cite{He2013} whereby, the 1st peak had the smallest height. Similar multi-peak natures is also being observed in the currently ravaging corona virus pandemic.\cite{Wangping2020} For example USA is already in its third peak, which is much higher than the previous two peaks; countries like Germany, Spain, France, England, etc. are suffering from a second infection wave, characterized by a rising second peak (\href{https://www.worldometers.info/coronavirus/}{Reference}). However, there are hardly any mathematical model available that can reproduce such behaviour, let alone predict it.

\begin{figure}[ht]
 \centering
 \includegraphics[width=3.3in,keepaspectratio=true]{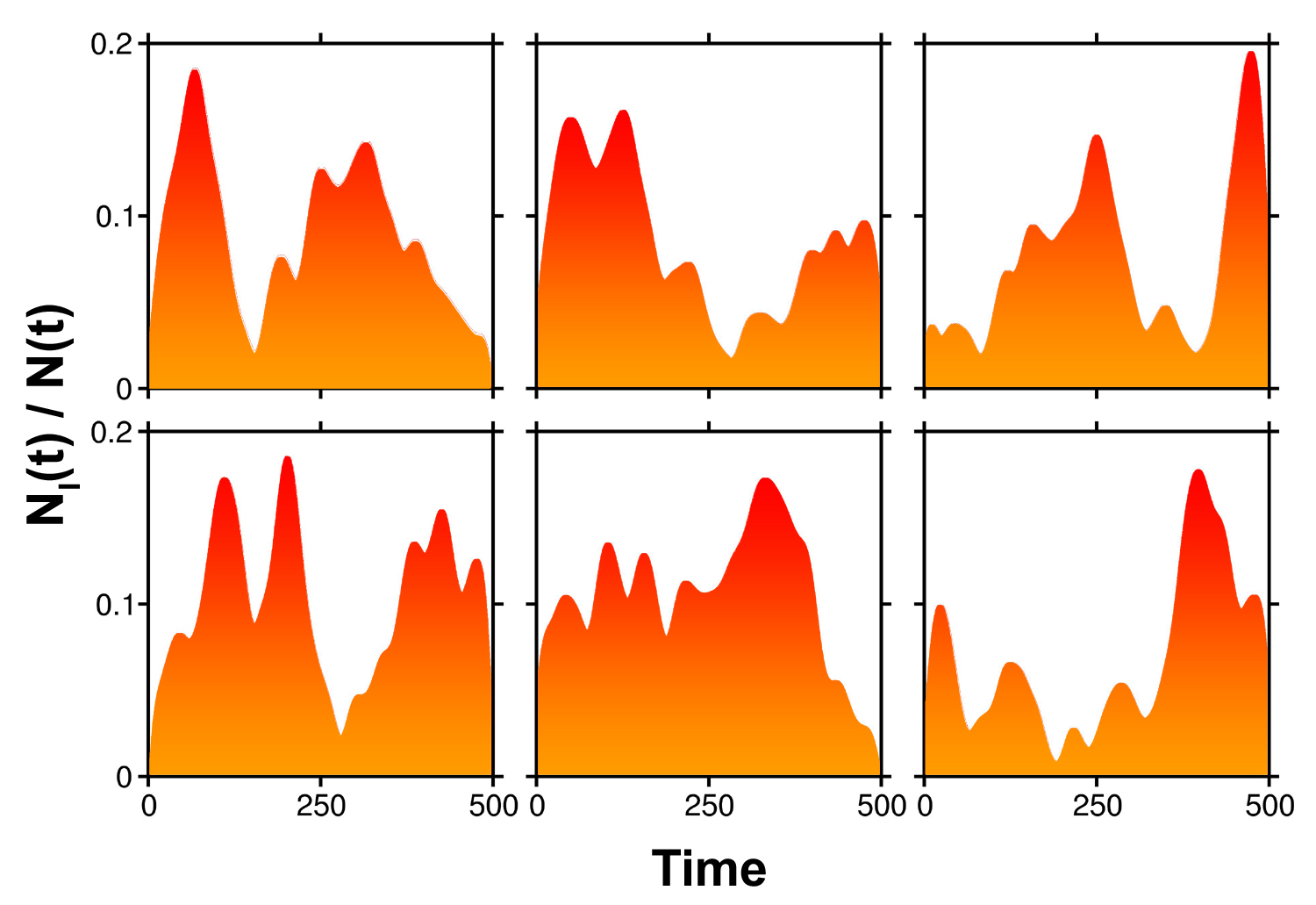}
 \caption{Time evolution of the fraction of population infected, showing multiple peaks in six separate simulations with different initial population configurations. The unit of time is explained in the caption of Fig. \ref{fig4}. The pattern of the peaks are different in each simulation. This mimics the different natures of evolution of disease in different localities or countries. There are three reasons primarily responsible for this multi-peak scenario: (a) distribution of inherent susceptibility, (b) distribution of external infectivity and (c) long-range migration of infected individuals.}
 \label{fig6}
\end{figure}

The multi-peak behaviour of a pandemic is not seen everywhere and there are certain factors that determine it. In our KMC-CA simulations, we implement the following factors that lead to the multi-wave nature of the infection curve (Fig. \ref{fig6}).

Population density in a country is spatially heterogeneous. In our earlier work, we have shown that with the increase in population density, the rate of infection increases.\cite{Mukherjee2020} Hence, in a region with high population density, the peak of infection is reached earlier. This, subsequently results in decay of the infection curve. This decay process might be further fuelled by national lockdown and increase in public consciousness. 

However, the restrictions need to be lifted after a certain period of time, whereby people can start travelling or migrating from one region to another. If, these migrants contain infected (asymptomatic or undetected) individuals, they can act as the nucleus in the new population of susceptibles, thus triggering a second pandemic wave. This is particularly true if the overall population density of the country is low. In case of higher density, the percolation of infection is facilitated by an easily available contact network of susceptible individuals. This could be a possible reason for the majorly single peak characteristics of the SARS-CoV-2 pandemic in India, in contrast to major European countries.

Another possibility is the change in seasons in the course of months through which the pandemic exists. Temperature and humidity may significantly alter the nature of the virus, thereby manipulating the infectivity indices of the infectives and asymptomatics. Consequently, the nature of the infection curve will also change.

As mentioned in the earlier section, we implement random seeding of infection in our simulated community to mimic the arrival of migrants. This factor is given by the transfer matric $T({\bf r'}\to {\bf r})$ in Eq.\ref{eq6}. We set up the initial configuration of our simulation for a 300 $\times$ 300 matrix with 1 \% covered by susceptibles and 0.005 \% covered by infectives. The movement of infected individuals are restrained by a quarantine probability ($P_Q$) of 0.9. For the sake of simplicity, we have not restrained the movements of susceptibles and asymptomatics ($P_{LD} = 0$). A seeding probability of 1 \% is used to simulate the migrant behaviour. Recovered people are given a 0.1 \% probability of getting reinfected by becoming susceptible. Due to the inherent stochastic nature of the simulation (which is true of a real world pandemic), each simulation run results in a different infection pattern, as shown in Fig. \ref{fig6}. However, all the simulations give multiple waves in the infection curve. It is clear, that the 1st peak does not always represent the highest wave. Since, it is impossible to predict the nature of the subsequent waves, the possibility remains, that a pandemic may present itself in a more dangerous form in the future. In fact, this seems to be true in the case of COVID-19.

\begin{figure}[ht]
 \centering
 \includegraphics[width=3.3in,keepaspectratio=true]{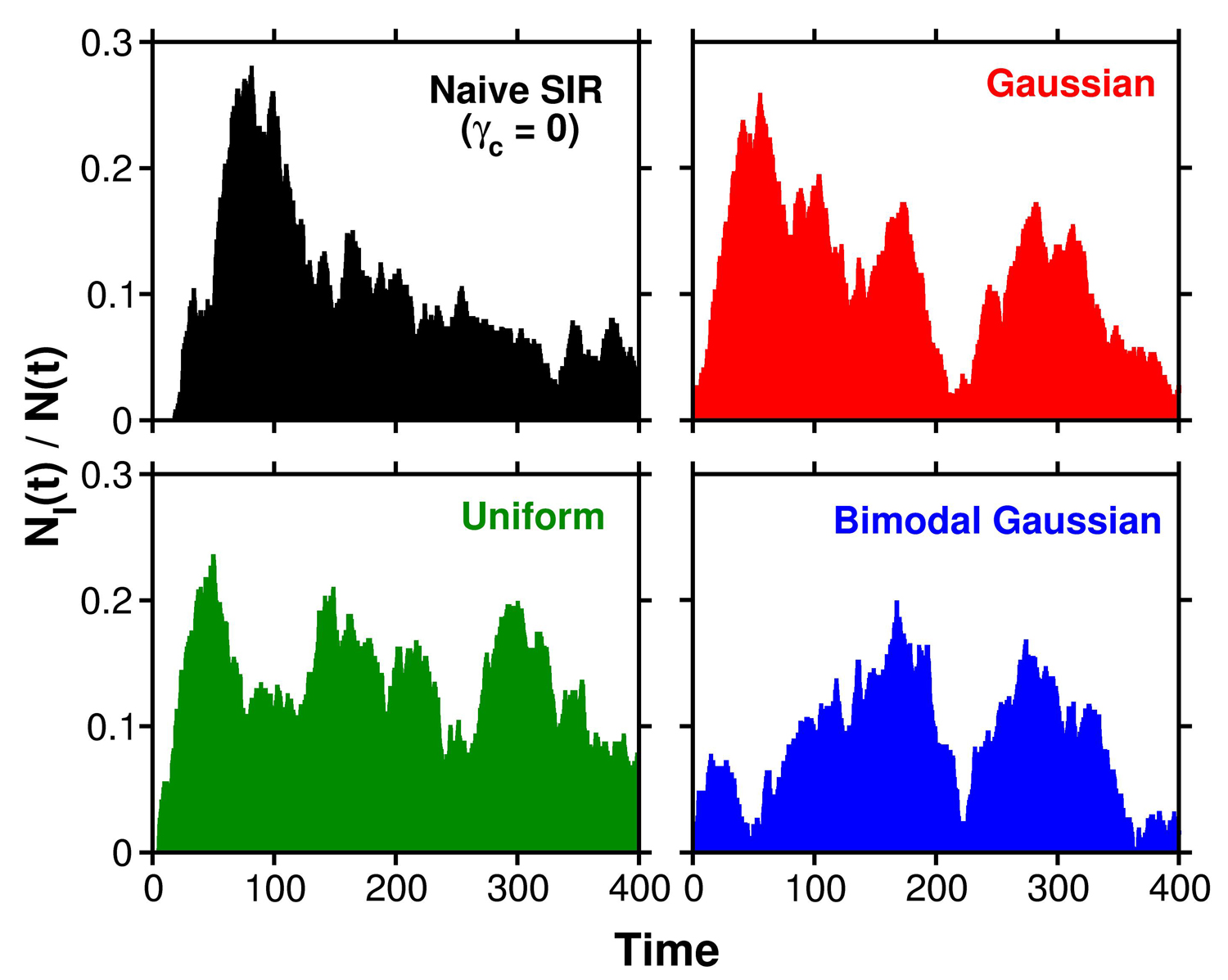}
 \caption{Temporal evolution of the fraction of infectives in a population under the effect of long-range migration. The unit of time is explained in the caption of Fig. \ref{fig4}. The susceptibility heterogeneity in the population is sampled in the population using three different distribution patterns: Gaussian (red), uniform (green) and bimodal Gaussian (blue) as shown in Fig. \ref{fig4}a. These are compared to the naive SIR model (black), which does not consider any distribution of susceptibility and susceptible individuals can get infected instantly in contact with infectives. Clearly, multiple peaks are observed when the distributions are considered, which shows that immunity heterogeneity plays an important role in the occurrence of the multiple infection waves. }
 \label{fig7}
\end{figure}

As shown in previous discussion, besides the long-range migration, susceptibility distributions play crucial role in the occurrence of the infection peaks. In Fig. \ref{fig7}, we show the infection curves in case of the different distributions introduced in Fig. \ref{fig}a. In the absence of the effect of any susceptibility distribution ($\gamma_c = 0$) (black), the multiple peaks are not clearly manifested. Whereas, in the other three cases (red: Gaussian, green: uniform, and blue: bimodal Gaussian) clear signatures of multiple peaks are observed. This shows that the occurrence of multiple infection waves is a combined effect of long range migration and susceptibility distribution.

It is interesting to note that the infection landscape can be considerably rugged with sharp falls and rises. These variations have their origin in the susceptibility and infectivity distributions we discussed earlier. The presence of these distributions clearly makes it a formidable problem even to venture a quantitative prediction of the progression.

\begin{figure}[ht]
 \centering
 \includegraphics[width=3.3in,keepaspectratio=true]{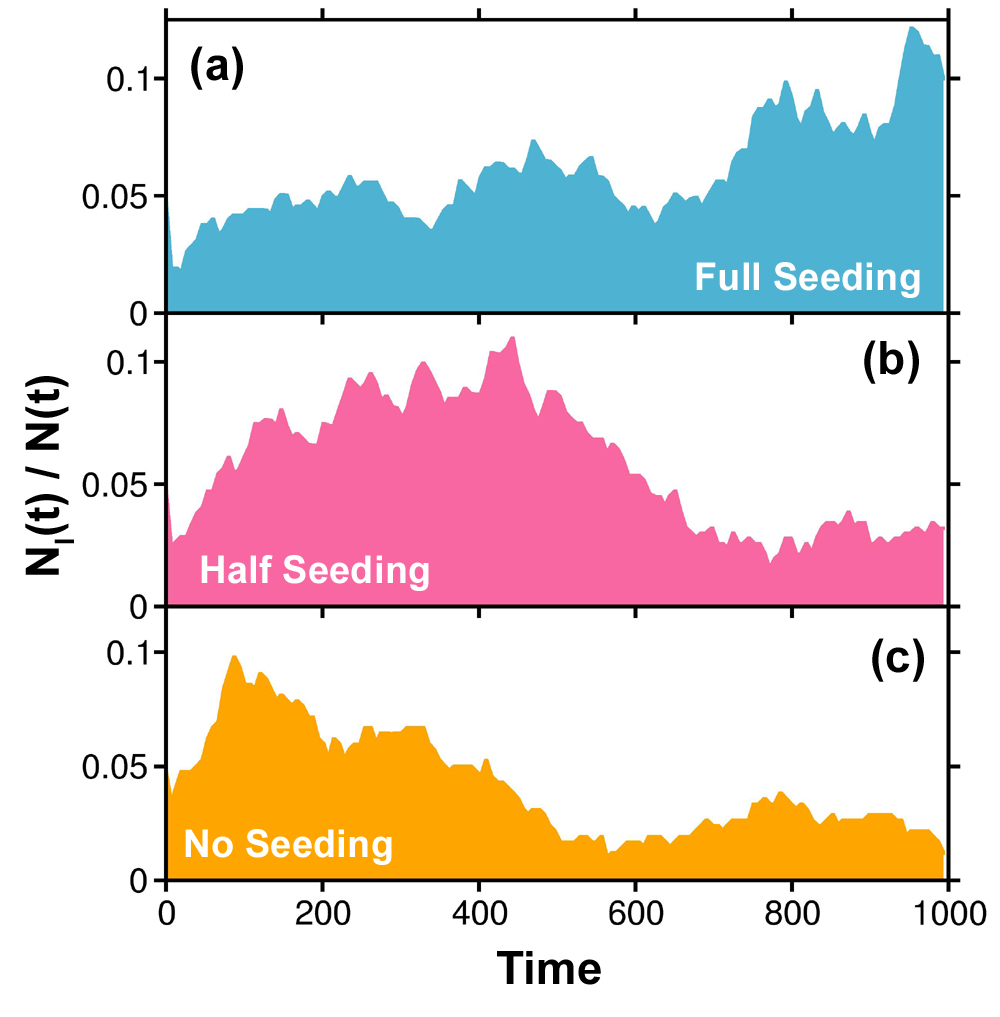}
 \caption{The effect of long-range migration on the time evolution of infection in a population. Such migration is enabled in our simulation via seeding of infected individuals at random places and at random times. The three graphs shown here represent the scenarios when infection seeding is (a) enabled throughout the simulation (full seeding), (b) enabled till half of the simulation time (half seeding), and (c) disabled (no seeding). The unit of time is explained in the caption of Fig. \ref{fig4}.}
 \label{fig8}
\end{figure}

The combined effect of these two factors particularly that of migration is further demonstrated in Fig. \ref{fig8}. Here we plot the fraction of infectives against time for the following three scenarios. Random infection seeding is (a) enabled throughout the simulation (full seeding). This represents the migration of infected or asymptomatic persons in the present population from the beginning, till the complete termination of the pandemic, (b) enabled till half of the total simulation time (half seeding), and (c) disabled, so that long-range migration does not add to the infection of the population.

This shows that random infection seeding (migration) results in a consistent rate of infection which slows down the decay of the curve. The infection starts to fall when migration stops (b). Even in absence of the migration (c), a second peak, though very small, can be observed. This results from susceptibility inhomogeneity. For full seeding of disease migration (a), several infection peaks are observed and very long simulations need to be run to obtain a complete decay in the number of infections.

The interaction between the seeding by migration and the presence of distributions can give rise to novel features like in Fig. \ref{fig8}. In the presence of large immunity in a population, the rise in infection initiated and forced by migration can undergo a slow decay without giving rise to a second peak. Thus the multiple peaks are a consequence not only of continuous seeding but also distributions. This could be the reason of the multiple surges of COVID-19 that we see in the European countries and also in the USA.

\section{Conclusion}
\label{sec6}

In this work we model the spatio-temporal evolution of a pandemic by generalization of the classical SIR model to include three important factors that strongly influence the progression of infectious diseases, yet have not been adequately addressed previously. These are (i) distribution of susceptibility, (ii) distribution of infectivity, and (iii) infection seeding via long range migration. We perform Kinetic Monte Carlo Cellular Automata (KMC-CA) simulations to solve the highly coupled and entangled master equations. Our analysis shows that the propagation of an infectious disease resembles a series of physical phenomena, which starts from a disease nucleation, from where the disease diffuses isotropically into the whole population. This results in a percolation network of the infection, causing an ultimate phase separation among the different compartments of the population.

Among many limitations of the naive SIR model, the absence of any treatment of the pre-existing heterogeneous distribution of population density and its disease propensity, long distance transfer of infection by migration, distributions of susceptibility and infectivity pose serious challenges when one attempts to apply the model to any  real world situation. Presence of the distribution alone makes a straightforward solution of the master equation virtually impossible. The effects of population density and infection density should be treated as separate entities that combine with susceptibility distribution to produce widely different patterns of infection in different regions and countries. In order to incorporate the distributions of age, activity, susceptibility, and infectivity several granular models have appeared. However, none of them could forecast regarding the multiple infectivity peaks and its origin.

Our model is a significant improvement over the classical SIR model.\cite{Kermack1927} Inclusion of the distributions of inherent susceptibility and external infectivity enables us to model a more realistic form of a pandemic. While the former defines the immunity of a susceptible person, the latter depends on several factors, such as hygiene of an infective, climate conditions, etc. A combination (product) of these two factors (Eq.\ref{eq2}) is used to determine the progression of the disease via human-to-human contact. We have also considered the movement/migration of disease vectors (infected, mainly asymptomatic individuals) from one place to another, via random seeding of an infective in our simulated society. This serves as an important agent that can trigger a pandemic in a non-affected region, which ultimately gives rise to infection waves, subsequent to the primary peak. We generalize the SIR model to include the non-local effects. We use of cellular automata to solve the nonlinear nonlocal equations.

From our analysis we find that the origin of the multiple infection peaks and the rugged infecton landscape is a combined consequence of all the three factors described above. Since the quantification of susceptibility and thus generation of a distribution is nontrivial, we use three model distributions in this work, namely (i) Gaussian, (ii) uniform, and (iii) bimodal Gaussian. We find that in absence of the distribution, the naive SIR model overestimates the extent infection in a society. This is true for all the three distribution patterns. Not only does this trigger erroneous mortality prediction, but also provides incorrect exstimates of the herd immunity threshold.

One aspect has become clear over the last few months – the progress of COVID-19 continues to thrive on a large number of factors that are hard to control. For example, an individual with low susceptibility may escape infection during the first wave, but fall victim during the time restrictions are eased. While we worry about reinfection, the former scenario could be of value in understanding the progression, because people with low susceptibility could become disease prone on long exposures, for example in in closed environments like offices and restaurants. 

The consideration of the critical infection parameter that we use to determine the propagation of an infection can also be used to model other cellular automata such as the propagation of a fire front.\cite{Almeida2011, XavierViegas1998, Encinas2007} For example, in the region of a wild fire incident, the effective dryness of the combustible material has a distribution determined by the dryness parameter and the immediate fire front has a distribution of hotness depending on the region, given by the hotness parameter. The fire front can only progress if the product of these two parameters attains a certain critical value. Hence, the model introduced in this work is a general cellular automata technique that can be applied to simulate other similar propagation phenomena.

\begin{acknowledgments}
B. Bagchi thanks SERB (DST, India) for the India National Science Chair Professorship and partial funding of the work, and also SERB (DST, India) for research funding. S. Mukherjee and S. Mondal thank IISc for the Research Associate fellowship.
\end{acknowledgments}

\bibliography{Pandemic}

\end{document}